\documentclass[lettersize,journal]{IEEEtran}
\usepackage{amsmath,amsfonts}
\usepackage{float}
\usepackage{array}
\usepackage{cite}
\usepackage[caption=false,font=normalsize,labelfont=sf,textfont=sf]{subfig}
\usepackage{textcomp}
\usepackage{stfloats}
\usepackage{url}
\usepackage{verbatim}
\usepackage{graphicx}
\usepackage{multirow}
\usepackage{amsmath}
\usepackage{amsfonts}
\usepackage{tabularx}
\usepackage[linesnumbered,ruled,vlined]{algorithm2e}
\usepackage{threeparttable}  % Add to preamble
\usepackage{multirow}
\usepackage{amsmath,amssymb}
\usepackage[
    colorlinks=true,
    citecolor=blue
]{hyperref}
\def\BibTeX{{\rm B\kern-.05em{\sc i\kern-.025em b}\kern-.08em
    T\kern-.1667em\lower.7ex\hbox{E}\kern-.125emX}}
\usepackage{balance}

\newcommand{\KwParam}[1]{\textbf{Parameters:} #1\par}
\begin{document}

% \title{HOMI: Ultra-Fast EdgeAI platform for Event Cameras}
\title{
{\scriptsize\textit{
This work has been submitted to the IEEE for possible publication. 
Copyright may be transferred without notice, after which this version 
may no longer be accessible.}}\\[-0.25em]
HOMI: Ultra-Fast EdgeAI platform for Event Cameras
}

\author{
\IEEEauthorblockN{
Shankaranarayanan H\textsuperscript{$\star$}\thanks{\textsuperscript{$\star$}These authors contributed equally to this work.}, 
Satyapreet Singh Yadav\textsuperscript{$\star$}, 
Adithya Krishna, 
Ajay Vikram P,
Mahesh Mehendale, \\
Chetan Singh Thakur \\
}
\IEEEauthorblockA{
Department of Electronic Systems Engineering, Indian Institute of Science, Bangalore, India 560012\\
}
}

\maketitle

\begin{abstract}
Event cameras offer significant advantages for edge robotics applications due to their asynchronous operation and sparse, event-driven output, making them well-suited for tasks requiring fast and efficient closed-loop control, such as gesture-based human-robot interaction. Despite this potential, existing event processing solutions remain limited, often lacking complete end-to-end implementations, exhibiting high latency, and insufficiently exploiting event data sparsity. In this paper, we present HOMI, an ultra-low latency, end-to-end edge AI platform comprising a Prophesee IMX636 event sensor chip with an Xilinx Zynq UltraScale+MPSoC FPGA chip, deploying an in-house developed AI accelerator. We have developed hardware-optimized pre-processing pipelines supporting both constant-time and constant-event modes for histogram accumulation, linear and exponential time surfaces. Our general-purpose implementation caters to both accuracy-driven and low-latency applications.
HOMI achieves 94\% accuracy on the DVS Gesture dataset as a use case when configured for high accuracy operation and provides a throughput of 1000 fps for low-latency configuration. The hardware-optimised pipeline maintains a compact memory footprint and utilises only 33\% of the available LUT resources on the FPGA, leaving ample headroom for further latency reduction, model parallelisation, multi-task deployments, or integration of more complex architectures.

\end{abstract}


\section{Introduction}
The demand for real-time perception in edge robotics, autonomous systems, and embedded platforms has increased interest in sensing and processing solutions operating under strict latency, and computational constraints. Rapid and reliable decision-making is essential in such applications, including gesture-based human-robot interaction, collision avoidance, and autonomous navigation \cite{trinh2018energy}. However, traditional frame-based vision systems are often unsuitable due to high data redundancy, bandwidth requirements, and processing latency. Event cameras, such as Dynamic Vision Sensors (DVS), offer a promising alternative by capturing asynchronous, sparse pixel-level changes with microsecond temporal resolution. This sensing paradigm enables fast, low-latency perception, making it particularly well-suited for real-time edge computing systems \cite{gallego2020event}.

Despite their promise, fully harnessing the potential of event cameras for edge computing requires more than sensor innovation; it necessitates a fundamental rethinking of the entire processing pipeline, from algorithm design to hardware acceleration, to meet the stringent demands of real-world robotic edge platforms \cite{sengupta2022embedded}. The key challenge lies not only in efficiently processing sparse event data but also in developing complete, end-to-end systems that integrate seamlessly with embedded and robotic platforms while satisfying the strict performance and latency requirements of time-sensitive applications \cite{wang2025towards}.

While numerous CPU/GPU-based solutions have demonstrated strong performance, they are often not designed for end-to-end deployment. Additionally, their high power consumption, large form factor, and significant computational requirements make them impractical for resource-constrained edge-embedded platforms \cite{sridharan2024ev}. Field Programmable Gate Array (FPGA) based designs offer a promising alternative, providing hardware configurability, parallelism, and low-power operation \cite{kryjak2024event, xu2020case}. However, most of the existing FPGA implementations fail to provide complete end-to-end integration \cite{yang2024evgnn,esda}, incur high processing latency \cite{amir2017low,linares2021dynamic}, demand substantial memory resources, and fail to exploit the inherent sparsity of event data \cite{tapiador2020event}.

We introduce HOMI, an ultra-fast, end-to-end EdgeAI platform built on the Xilinx Zynq UltraScale+ MPSoC FPGA and integrated with the IMX636 event sensor, all within a compact 10 cm $\times$ 5 cm form factor. To demonstrate its capabilities, we evaluate it on the DVS Gesture dataset, achieving latency of 1 ms (equivalent to 1000 fps) with a modest trade-off in accuracy, while using only 33\% of the available LUTs. In HOMI, we have deployed the in-house developed RAMAN accelerator that leverages both structured and unstructured sparsity to optimise performance and efficiency \cite{krishna2024raman}. Furthermore, sub-millisecond latency can be achieved by leveraging parallel computing blocks, given that only 33\% of the resources are utilised, allowing room for deploying more complex models and supporting multi-task execution. These results establish a new benchmark for real-time classification pipelines tailored for edge computing in robotic platforms. The key contributions of this work are as follows:
\begin{itemize}
\item\textbf{Design of the HOMI, an end-to-end ultra-fast FPGA-based EdgeAI platform}, that interfaces a Prophesee IMX636 event sensor \cite{propheseeIMX636} chip with Xilinx Zynq UltraScale+MPSoC FPGA chip over a high-speed Mobile Industry Processor Interface (MIPI) along with the CNN Accelerator through a robust and hardware-friendly pre-processing block on the FPGA Programmable Logic (PL) fabric, resulting in low-latency compute, which in turn facilitates its deployment for real-time high-speed applications. Besides inference tasks, HOMI can also be leveraged for real-time data acquisition and visualisation applications, bringing a paradigm shift for the next-gen event-based systems.

\item Proposed a novel Shift-based Linear Time Surface (SLTS) and Shift-based Exponential Time Surface (SETS) representations, which are hardware-efficient alternatives to traditional time surfaces that eliminate the need for memory-intensive LUTs. By leveraging simple shift operations, SLTS and SETS significantly reduce computational complexity and memory overhead, enabling real-time, CNN-compatible event frame generation.
\item Developed a low-latency, highly customizable and novel pre-processing block capable of generating different event representations in constant-time and constant-event modes for resolutions up to High-Definition (HD) $1280 \times 720$. It supports binary frames, histogram accumulation, SLTS and SETS.

\item Collected an in-house DVS Gesture dataset at a resolution of $1280 \times 720$ from five participants performing 11 standard gesture classes, consistent with the original DVS Gesture dataset \cite{amir2017low}. This high-resolution dataset was used to validate our complete end-to-end pipeline and will be made publicly available.
\end{itemize}

The remainder of this paper is organised as follows. Section II reviews event cameras and related FPGA-based classification pipelines. Section III outlines the methodology, system architecture, data collection, and training pipeline. Section IV presents the experiments and results. Section V concludes the paper and discusses future work.

\section{Background and Related Work}

Event cameras depart from traditional frame-based imaging by mimicking the spike-driven, asynchronous processing of biological retinas. Each pixel operates independently, generating an event when the logarithmic change in brightness exceeds a set threshold, encoding the pixel’s spatial coordinates $(x,y)$, a $\mu$s-precision timestamp $t$, and polarity $p$ indicating intensity increase ($+ 1$) or decrease ($- 1$) \cite{lichtsteiner200564x64, lichtsteiner2006128}. This event-driven sensing enables ultra-low latency, high temporal resolution, and low power consumption, as data is produced only in response to scene changes. The logarithmic response ensures a wide dynamic range ($>$ 120 $dB$) \cite{lichtsteiner2008128}, while the asynchronous operation inherently eliminates motion blur. These properties make event cameras highly suitable for robotic applications that demand fast, efficient, and robust perception in dynamic, low-power, and resource-constrained environments.

\par One of the key challenges in developing end-to-end event-based systems lies in the efficient interfacing of event cameras. Earlier implementations \cite{amir2017low,tapiador2020event,linares2021dynamic} have relied on Universal Serial Bus (USB) interfaces to stream data from off-the-shelf DVS cameras in Address Event Representation (AER) format. However, the limited bandwidth of USB significantly constrains system performance, introducing latency and limiting scalability. To improve throughput, ColibriUAV \cite{colibriuav} adopted a Synchronous AER (SAER) interface. While this approach improved the throughput for the simple inference task, it did not fully overcome the bandwidth limitations inherent to AER-based protocols. To address these bottlenecks, Swifteagle \cite{swifteagle} which is primarily designed as a data acquisition system, interfaced the DVS sensor directly with the K26 System on Module (SOM) via a MIPI CSI-2 lane. The MIPI RX subsystem supports data rates up to 1.6 Gbps \cite{xilinx_mipi}, sufficient to accommodate the peak bandwidth requirements (~1.5 Gbps) of modern high-speed sensors like the Prophesee GenX320 and IMX636 (CCAM5).  Despite this advancement, the use of the EVT 2.1 encoding format \cite{prophesee_evt21} introduces inefficiencies—both in terms of data compaction and I/O bandwidth utilization—compared to the more recent EVT 3.0 format \cite{prophesee_evt30}. These inefficiencies result in under utilization of the available MIPI bandwidth and introduce additional overhead in downstream decoding. Moreover, existing implementations typically use low-resolution DVS sensors, limiting the spatial granularity of captured events. Although recent developments have introduced high-resolution sensors—offering resolutions up to $1280 \times 720$ \cite{gallego2020event}—integrating such sensors poses further challenges in terms of data throughput, memory bandwidth, form-factor and real-time processing requirements. As a result, despite improvements in interfacing mechanisms, current systems remain constrained by format inefficiencies, resolution limitations, and processing bottlenecks, highlighting the need for a holistic co-design approach across sensor interface, data encoding, and hardware acceleration.

CNNs are preferred for general-purpose event-data processing on FPGAs due to their compatibility with standard accelerators, ease of configuration, and widespread user familiarity \cite{mittal2020survey}. In contrast, Spiking Neural Networks (SNNs), while energy-efficient, require specialised hardware such as Loihi \cite{davies2018loihi} or SpiNNaker \cite{furber2014spinnaker} to achieve competitive performance, limiting their practical adoption due to the need for dedicated neuromorphic platforms \cite{bouvier2019spiking}.

The sparse nature of event-based data presents a compelling opportunity to reduce computational complexity and latency in CNNs by exploiting activation and weight sparsity. While numerous FPGA-based accelerators have been proposed for general-purpose CNN inference, this work specifically focuses on those integrated into end-to-end event-based systems, along with a few that leverage sparsity for efficient inference. Notably, NullHop \cite{nullhop} introduces a zero-skipping architecture that exploits activation sparsity to reduce memory and compute demands. However, it does not support sparse weight processing or dynamic pruning techniques. A complete system implementation based on NullHop is presented in \cite{linares2021dynamic}, where a compact CNN model achieves 160 fps, demonstrating the feasibility of deploying sparsity-aware accelerators in real-time event pipelines. ESDA \cite{esda}, on the other hand, adopts a composable sparse dataflow architecture that dynamically exploits activation sparsity, achieving a throughput of 839 fps on MobileNetV2 for the DvsGesture dataset. Despite its high throughput, ESDA incurs significant resource usage, which limits its scalability and reconfigurability. The more recent RAMAN accelerator \cite{krishna2024raman} addresses both activation and weight sparsity within a reconfigurable framework, while also incorporating quantisation support for latency reduction. It offers improved resource efficiency compared to ESDA, making it a practical candidate for deployment in compact, end-to-end event-based inference platforms.

While FPGA-based CNN pipelines for event data show promise, efficient feature generation remains a challenge, affecting system latency and resource efficiency. Histogram accumulation methods are simple to implement \cite{bonazzifpgadrone} but discard temporal information by merely accumulating events in fixed time window, thereby limiting performance. Few implementations introduce normalization steps \cite{linares2021dynamic,bonazzifpgadrone} that add processing overhead, reducing real-time performance.

To overcome the limitations of histogram accumulation and capture rich spatio-temporal patterns, the neuromorphic community introduced time surfaces. LTS apply a linear decay, giving equal weight to all recent events within a time window, while ETS assigns higher importance to more recent events through exponential decay \cite{lagorce2016hots, sironi2018hats}. By computing time surfaces over a fixed number of events, these surfaces naturally achieve speed invariance and offer rich spatio-temporal features. While LTS is straightforward to implement, ETS is resource-heavy on FPGAs as it demands additional LUTs to store precomputed exponential values, leading to increased memory and power consumption. ETS designs have been reported to use approximately 5.4× more LUTs compared to histogram-based approaches at $1280 \times 720$ resolution \cite{blachut2023high}.

\section{Methodology}

\begin{figure*}[!t]
    \centering
    \includegraphics[width=\textwidth]{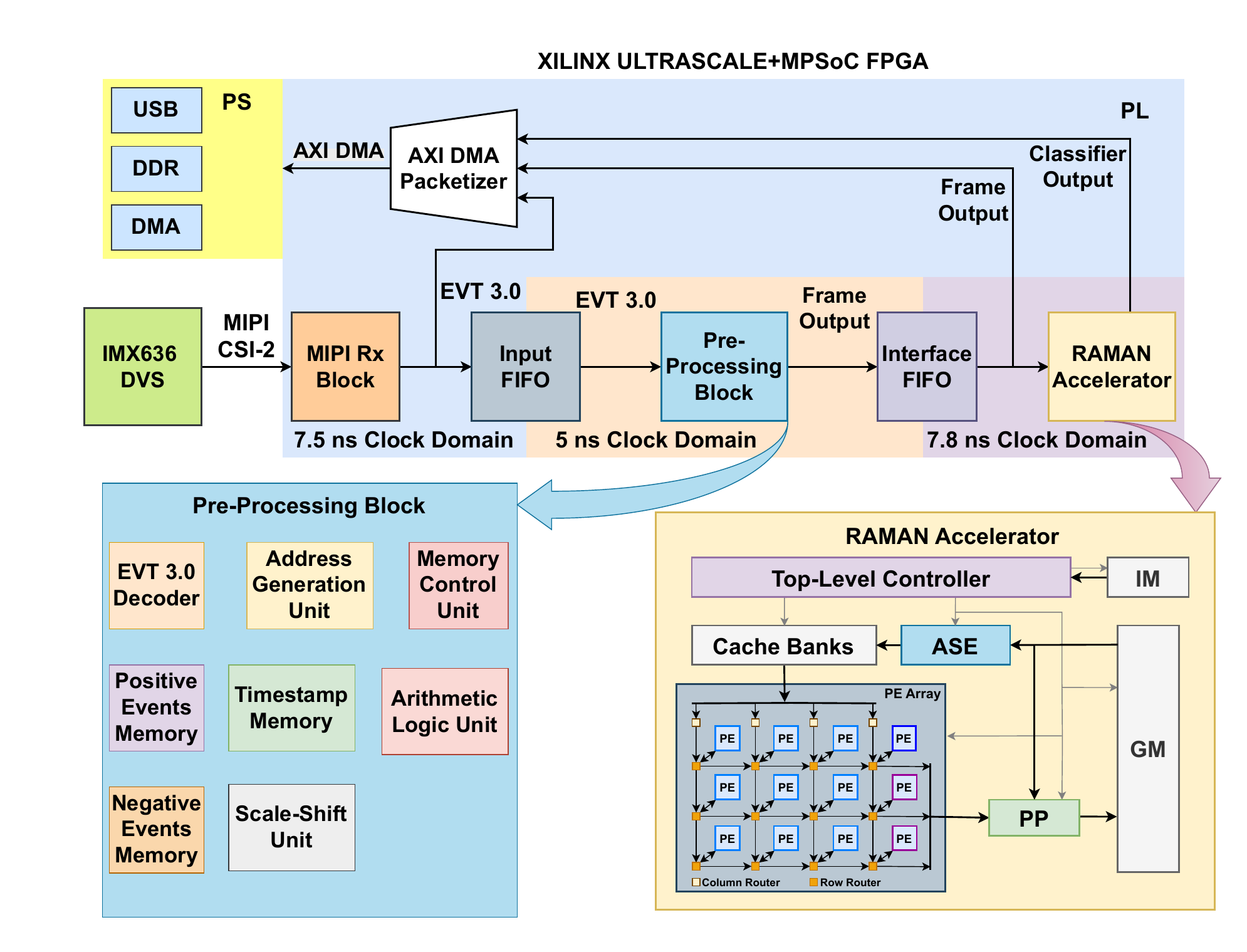}
    \caption{End-to-end flow of the HOMI comprising of the IMX636 DVS Sensor, MIPI RX Block, Pre-Processing Block, Interface FIFO and the RAMAN Accelerator \cite{krishna2024raman}. The pre-processing block takes the inputs in the native EVT 3.0 format and generates the event representation on basis of application requirements. Once the representation is generated, it is fed into the accelerator via the interface unit. The AXI DMA packetizer module acts as a bridge between the processing system and programmable logic on the FPGA. Either of the raw events, or generated frames or the classifier output is sent to the processing system through the means of DMA, which can be later stored in the DDR memory for acquisition or sent out via the USB interface for visualization. The RAMAN Accelerator comprises of a global memory for storing the parameters and the activations, an Instruction Memory (IM) for storing the layer information, a run-time activation sparsity engine (ASE), cache banks for temporal data reuse, an array of PEs to do the MAC operations followed by a post-processing block which writes back the output to the global memory (GM).}
    \label{endtoend_flow}
\end{figure*}

Fig. \ref{endtoend_flow} presents the end-to-end flow of HOMI. For this work, we have deployed the RAMAN accelerator for its advantages in latency reduction and scalability. In this section, we explain the functionality and hardware implementation of the various modules involved and their interfaces.
\subsection{Sensor and PS-PL Interface}
For this work, we have used an Xilinx Ultrascale+MPSoC FPGA, which comprises the Processing System (PS) and the Programmable Logic (PL). As depicted in Fig. \ref{endtoend_flow}, the design can be divided into three clock domains. The sensor interface and Advanced eXtensible Interface (AXI)  Direct Memory Access (DMA) Packetizer modules connected to the PS work at a clock period of 7.5ns. The pre-processing logic and the RAMAN accelerator operate with clock periods of 5 ns and 7.8 ns, respectively. The MIPI RX Block receives the event stream from the DVS Sensor in the EVT 3.0 format, which is carried into the blocks in AXI4-Stream format. The required configuration data for the sensor is sent to its memory-mapped registers upon processing system boot-up. The AXI DMA Packetizer module bridges the PL and the PS parts of the FPGA. With the help of DMA, the required data from the PL side is transmitted to the PS. The data can be the raw events, generated frames, final prediction values or both, based on the inputs to the select line of the output transmission multiplexer. The data transmitted via DMA to PS is stored in the Double Data Rate (DDR) memory connected to the PS via the DDR controller. Upon request, the PS can stream the data stored in the DDR memory to the host PC via the high-speed USB link.

\subsection{Motivation behind Multiple Clock Domains}

One of the primary reasons for setting up multiple clock domains is the event stream format. EVT 3.0 consists of a vectorized input format, which, in turn, eliminates the need for transferring redundant data. For example, depending on the sensor type \cite{prophesee_genx320,propheseeIMX636}, the event stream may be organised into 10 or 40 banks of 32 pixels each, respectively. Suppose two or more events of the same polarity are triggered in different pixel banks. In that case, the sensor will send individual X addresses of the pixels; however, if they are triggered in the same pixel bank, then the sensor will first send the bank offset followed by a 32-bit vector in the chunks of 12, 12 and 8 bits (along with the event type which overall packs up to 16 bits for each case) where in the presence of an event, the bit is set to 1 else it is 0. In this way, if 32 events of the same polarity happened in the same pixel bank, then instead of transmitting 32 times the different values of X accounting for the overall transfer of 64 bytes, the whole information is transmitted with 8 bytes. This encoding logic helps reduce the system's overall data and IO bandwidth, but complexity arises when event-based representations need to be generated.

\par As described in \cite{swifteagle}, the interface and acquisition platform operates at a clock of 10ns, and because there is no pixel-level vectorization support in the EVT 2.1 format, it is relatively straightforward to generate the frames for different event representations on-the-fly depending upon the received event stream values. However, in the case of EVT 3.0 format, if a pixel-level vectorized input is encountered, the decoding logic has to account for each pixel in the vector before moving on to the following input. Along with this, the accelerator is being operated at 128MHz, and if the pre-processing block also operates at the same frequency, then there is a possibility of missing the events. For these reasons, we operate our pre-processing logic at a relatively faster clock of 5ns compared to the sensor clock to account for the vectorized inputs and inference time by the accelerator to ensure that no events are missed. This mechanism avoids complex algorithms and techniques to drop events if the Input First-In-First-Out (FIFO) buffer is filled \cite{blachut2023high}. 

\begin{figure*}[!t]
    \centering
    \includegraphics[width=\textwidth]{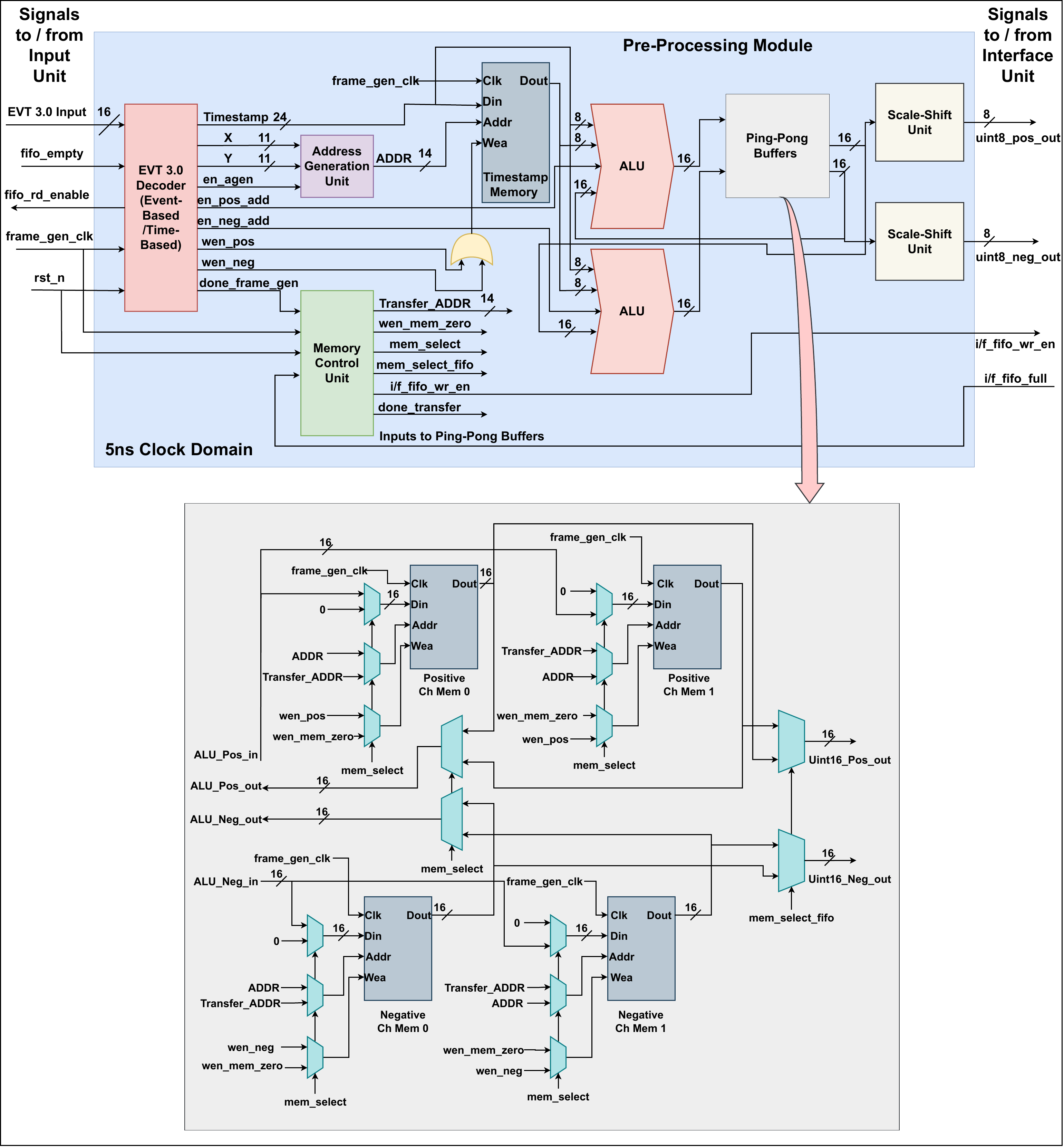}
    \caption{Overview of the Pre-Processing Module. It contains the decoder, address generation unit, ALUs, memory control unit to manage the ping-pong buffers and scale-shift unit for quantizing the 16-bit output to 8-bit to make it compatible with the adapted accelerator. A more detailed block diagram of the ping-pong buffers is depicted with the appropriate data and control signals to each of the buffers.}
    \label{pre_process_overview}
\end{figure*}

\subsection{Pre-Processing Module}
\par Fig. \ref{pre_process_overview} presents the overview of the pre-processing module. It comprises constant event-based and constant time-based control units for decoding the incoming EVT 3.0 data, an address generation unit, a memory control unit, memories for storing time-stamp, positive and negative event representations, an arithmetic logic unit (ALU) for generating different event representations, followed by a shift-scale block which maps them to an unsigned 8-bit format. The output from the pre-processing block is forwarded to the interface unit for further processing.

\begin{figure}[!t]
  \centering
  \includegraphics[width=0.8\columnwidth]{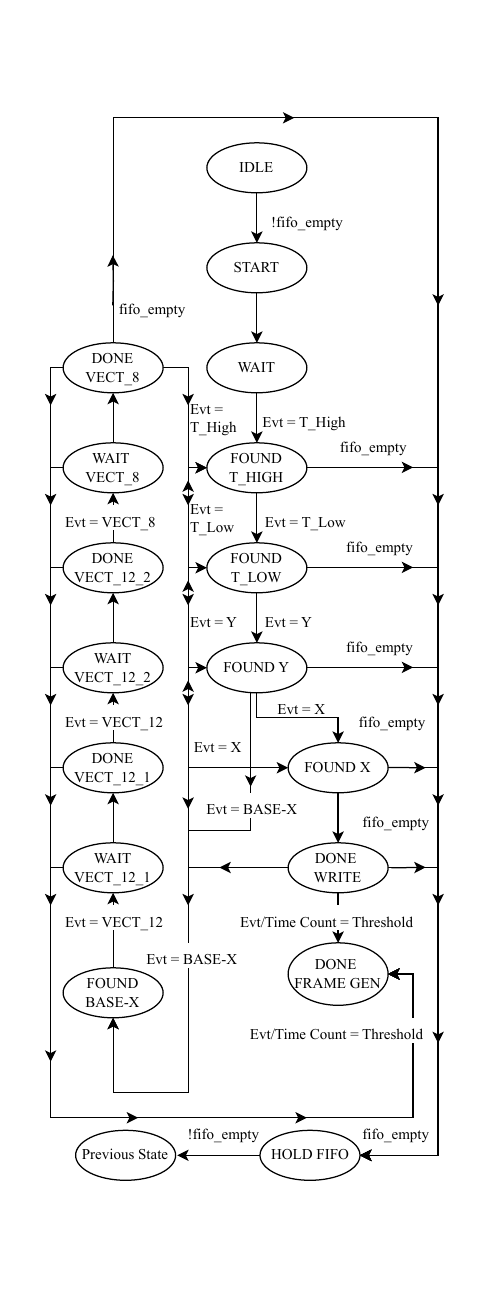} % Use width slightly under 3.5in
  \caption{Flowchart of the Event-based and Time-based Control Units.}
  \label{controller_flow_chart}
\end{figure}
\subsubsection{Event-Based and Time-Based Control Units}
\par Fig. \ref{controller_flow_chart} depicts the flowchart for the event-based and time-based control units. The controller flow is the same for both cases, except that, in the case of event-based, the flow keeps track of the number of events, which results in variable accumulation time depending on the scene dynamics. For the time-based case, synchronous sampling of frames is done at an integer multiple of the pre-processing clock. For the current use case, we are spatially downsampling the frames from $1280 \times 720$ to $128 \times 128$, which results in 16,384 locations for the memories and two buffers per channel for simultaneous acquisition and transfer. Hence, as a lower bound for this case, the number of events to be acquired is 16,384. In the case of time-based, the maximum acquisition fps is set to 12,200 fps, as this many cycles are required to transfer the generated frame data to the other stages, thereby avoiding any issues about data overlap. Also, as the pre-processing block operates at a relatively higher frequency than the sensor interface side, the controller goes into the hold state if the input FIFO is empty and returns to the previous state, once the FIFO empty flag is deasserted and continues with the execution flow.

\par To handle the case of vectorized inputs efficiently, separate sub-controllers are designed such that instead of traversing through each vector bit sequentially, they consider only the valid bit locations for updation, thereby reducing the processing time for accumulation.

\par Once the event-based or the time-based accumulation is complete, the controller generates a done signal, which acts as a trigger for the memory control unit to perform further processing. 

\subsubsection{Address Generation Unit}\label{subsec:addr_gen}
\par The IMX636 sensor provides a spatial resolution of $1280 \times 720$, which is significantly higher than required for the DVS Gesture use case, as the original dataset was collected using a DVS sensor with a resolution of $128 \times 128$ \cite{amir2017low}. Hence, for this case, the input coordinates $(x_{\text{in}}, y_{\text{in}})$  from a higher resolution grid ($1280 \times 720$) are linearly mapped to output coordinates $(x_{\text{out}}, y_{\text{out}})$ in a lower resolution grid ($128 \times 128$), thereby removing the need for generating a high resolution intermediate frame. To convert high-resolution input coordinates $(x_{\text{in}}, y_{\text{in}})$ to output coordinates $(x_{\text{out}}, y_{\text{out}})$ in a downsampled 2D grid, we use a linear LUT-based mapping:

\vspace{-0.6cm}

\begin{align}
x_{\text{out}} &= m_x[x_{\text{in}}] \cdot x_{\text{in}} + b_x[x_{\text{in}}] \\
y_{\text{out}} &= m_y[y_{\text{in}}] \cdot y_{\text{in}} + b_y[y_{\text{in}}]
\end{align}

Since the slope values $m_x[i], m_y[i] \in \{0, 1\}$, the multiplications can be implemented using 2-to-1 multiplexers. However, because the inputs are in Q16 format, they must be shifted before use.

The optimized form becomes:

\begin{align}
x_{\text{out}} &= 
\begin{cases}
(x_{\text{in}} \gg 16) + b_x[x_{\text{in}}], & \text{if } m_x[x_{\text{in}}] = 1 \\
b_x[x_{\text{in}}], & \text{if } m_x[x_{\text{in}}] = 0
\end{cases} \\
y_{\text{out}} &= 
\begin{cases}
(y_{\text{in}} \gg 16) + b_y[y_{\text{in}}], & \text{if } m_y[y_{\text{in}}] = 1 \\
b_y[y_{\text{in}}], & \text{if } m_y[y_{\text{in}}] = 0
\end{cases}
\end{align}

\noindent
Here, \texttt{$\gg$ 16} denotes a logical right shift by 16 bits to convert from Q16 fixed-point to integer representation. This enables an efficient hardware implementation using only adders, shifts, and multiplexers. Finally, the memory address for storing or accessing the corresponding data in the Block Random Access Memory (BRAM) is computed as:

\begin{equation}
    \text{ADDR} = (y_{\text{out}} \ll 7) + x_{\text{out}}
\end{equation}

\noindent where $\ll 7$ denotes a bitwise left shift by 7 positions, equivalent to multiplying $y_{\text{out}}$ by 128 (i.e., the output width), thus preserving row-major memory layout along with reducing the critical path by avoiding Digital Signal Processors (DSPs) for multiplication.

\subsubsection{Representation Memories}
\par For storing the event representation, we instantiate BRAM IPs with the following features for this particular use case:

\begin{table}[h]
\centering
\caption{BRAM Configuration for event representations}
\begin{tabular}{|l|c|}
\hline
\textbf{Parameter} & \textbf{Value} \\
\hline
Depth & 16384 \\
Width (bits) & 16 \\
Read Latency (cycles) & 1 \\
No. of Ports & 2 \\
\hline
\end{tabular}
\label{tab:bram_config}
\end{table}
\par As we intend to operate in a ping-pong fashion, there is another set of BRAMs, resulting in a total of two BRAMs for each channel as shown in Fig. \ref{pre_process_overview}. Additionally, for storing the timestamp information of 24 bits, we have a BRAM of the same depth and read latency as described before. All the BRAMs are operated in a read-first mode, wherein in the first cycle, an address is sent, and in the second cycle, the data is read, modified and written back to the corresponding memory location targeted by the address. As the event representations depend on the previous values stored in the representation memory, it is important to ensure that the previous values are flushed out before the beginning of a new accumulation cycle, which is handled by the memory control unit. 
\subsubsection{Memory Control Unit (MCU)}
\par Along with transferring the event representation for inference and display, the memory control unit ensures resetting the memory by overwriting the previous data with 0 in the representation memory. Upon receiving the input trigger as a done signal from the frame generation unit, the MCU switches the representation memory for acquisition whilst transferring the acquired representation data to the FIFOs located at the interface unit. As shown in Fig. \ref{pre_process_overview}, multiplexers are kept at the inputs and outputs of representation memories, for operating in the ping-pong mode based on the control inputs from the MCU. At present, for display, we are transferring only one type of representation (i.e., either positive or negative channel); however, for inference, we can choose between feeding only single-channel data or multi-channel data, depending upon the model requirements. If either of the FIFOs becomes full, the controller goes into the hold state until the FIFO full flag is de-asserted. 
\subsubsection{Arithmetic Logic Unit (ALU)}
\par The Arithmetic Logic Unit (ALU) plays the role of updating the event representations during the accumulation phase. Separate ALUs are being utilized for positive and negative channels. Depending upon the type of representation needed, the ALU performs the corresponding operation. In the case of the binary frame, the ALU assigns the value of 255 to the downsampled memory location where the intensity change in the pixel was detected. In the case of histogram accumulation, for every positive and negative change in the pixel, the downsampled memory location is updated by a value of 1 till the accumulation phase is over. The equations for the same are described below:
\begin{align}
Mem[\text{ADDR}] &\leftarrow Mem[\text{ADDR}] + 1 \label{eq:accumulation_flat} \\
Mem[\text{ADDR}] &\leftarrow 
\begin{cases}
255, & \text{if } P = 1 \\
Mem[\text{ADDR}], & \text{otherwise}
\end{cases}
\label{eq:update_p_flat}
\end{align}

\noindent
Equation~\eqref{eq:accumulation_flat} performs an accumulation by 1, while Equation~\eqref{eq:update_p_flat} overrides the value to 255 if the change is detected at the corresponding memory location. Along with histogram accumulation and binary representation, we can also perform exponential and linear time surfaces by modelling the decay in a shift-based manner, which is very efficient from the hardware implementation perspective.

\subsubsection{Shift-Based Time Surfaces}
To efficiently model temporal decay without relying on floating-point operations, we approximate the decay function using bit-shift operations. For each event at spatial location \((x, y)\) with timestamp \(t_i\), the time difference \(\Delta t = t_i - T_{\text{last}}(x, y)\) is computed relative to the last event at that pixel. The decay term is then estimated using a right-shift operation, as shown below for SETS and SLTS for decay parameter $\tau \in \mathbb{Z}^+$:

\begin{equation}
\begin{aligned}
\text{decay}_{\text{SETS}} &= 2^{-\left( \Delta t \gg \tau \right)} \\
\text{decay}_{\text{SLTS}} &= \Delta t \gg \tau
\end{aligned}
\end{equation}

\begin{figure}[!t]
  \centering
  \includegraphics[width=0.9\columnwidth]{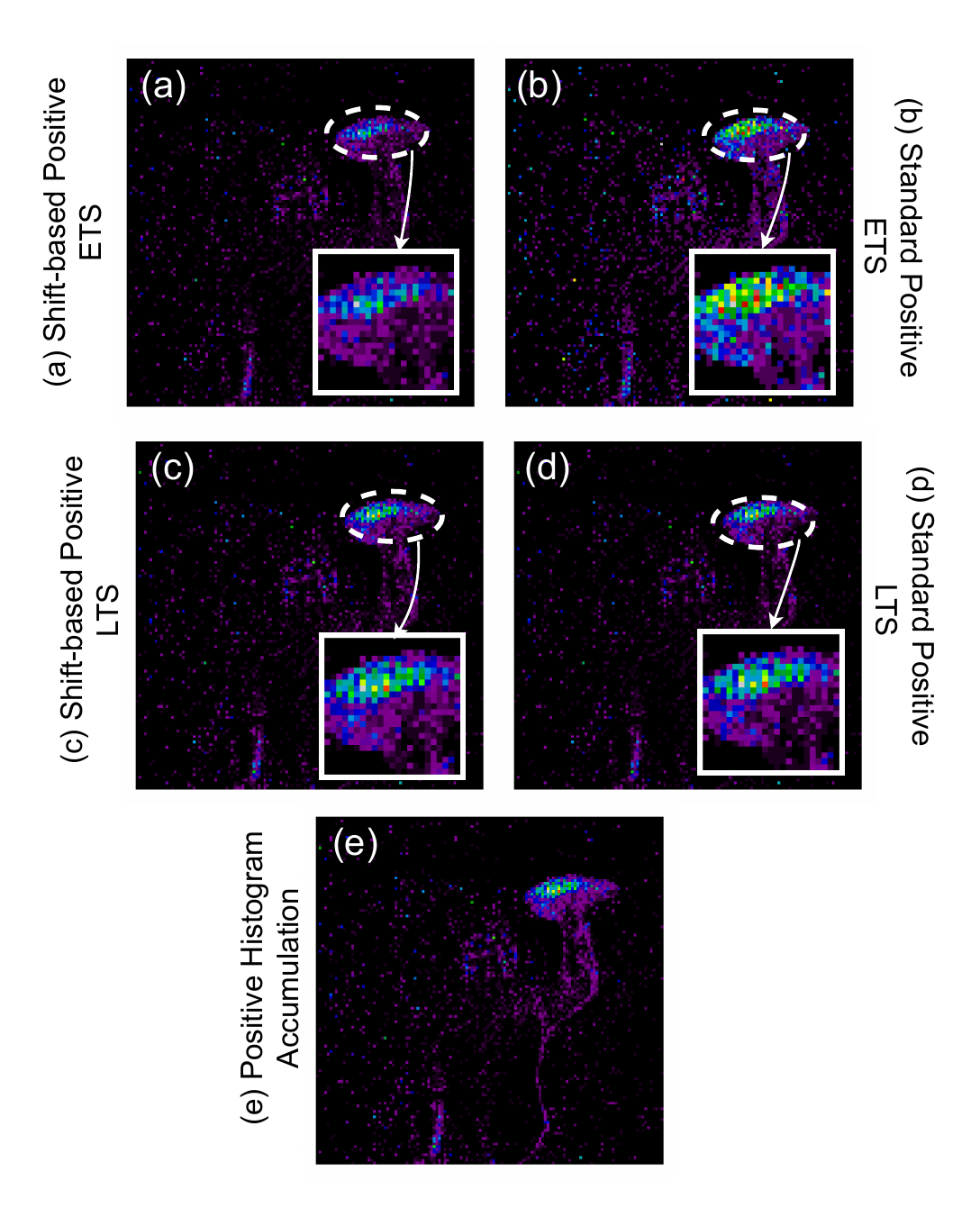}
    \caption{Comparison of shift-based and standard implementations of ETS, LTS, and histogram accumulation for positive polarities, using the \textit{Left Hand Wave} gesture from the DVS-Gesture dataset with 20K events. Subfigures (a)–(b) show ETS variants; (c)–(d) illustrate LTS variants; and (e) depicts histogram accumulation. The shift-based implementations retain the essential features of their standard counterparts, as demonstrated by the similarity observed in the zoomed-in profiles.}
  \label{time_surfaces}
\end{figure}

This approximation replaces exponentiation with efficient integer shifts and additions, reducing computational complexity and memory requirements. As seen from Fig. \ref{time_surfaces}, the bit-shift approximation preserves the fundamental characteristics of exponential decay, including its monotonic behavior, while introducing a controlled level of quantization. The relationship between the proposed approximation and the standard exponential function for a given $\tau \in \mathbb{Z}^+$ is given by
\begin{equation}
2^{-\left( \Delta t \gg \tau \right)} \approx e^{-\frac{\Delta t \cdot \ln(2)}{2^{\tau}}}
\end{equation}

\begin{algorithm}[t]
\caption{Shift-based Time Surface Update}
\label{alg:shift_time_surface}
\KwIn{Event at pixel $(x, y)$ with timestamp $t$}
\KwParam{$\tau \in \mathbb{Z}^+$ (integer decay parameter, e.g., $\tau = 16$)}
\KwParam{$T_{last}(x, y)$: last timestamp at $(x, y)$}
\KwParam{$S(x, y)$: time surface value at $(x, y)$}

$\Delta t \gets t - T_{\text{last}}(x, y)$ \tcp*[r]{Time since last event}
$shift \gets \Delta t \gg \tau$ \tcp*[r]{Shift-based decay}

\textbf{SETS Update:} \;
\If{$shift < 16$}{
    $S(x, y) \gets 1 + (S(x, y) \gg shift)$\;
}
\Else{
    $S(x, y) \gets 1$\;
}
$T_{last}(x, y) \gets t$\;

\textbf{SLTS Update:} \;
$S(x, y) \gets 1 + \max(0, S(x, y) - shift)$\;
$T_{last}(x, y) \gets t$\;
\end{algorithm}

\par Algorithm~\ref{alg:shift_time_surface} describes the flow for the shift-based time surface update. In the hardware, for generating the decay term, instead of shifting, we can consider the difference of the upper 8-bit from the 24-bit values of the present and past timestamp values. In the case of the SETS, if the decay is less than 16, then this decay shifts previous data read from memory; else, the memory is overwritten with value 1. Similarly, in the case of SLTS, the decay term can be subtracted from the previous data read in the memory. If the decay factor exceeds the previous data, the memory is overwritten with the value 1. Given that the timestamp value is 24-bit, there can be instances where the past value of the timestamp may be greater than the present value, suggesting that the counter has reset. In this regard, we compare the timestamp value with 0 to avoid this reset issue. As the BRAM used for storing the timestamp values is also used in read-first mode, the write enable signals used during the accumulation time for the representation memories can be used as the write enable signals for the timestamp memory to store the present timestamp values after the compute. Thus, at the hardware level implementation, the equations for SETS and SLTS can be described as follows:

{\small
\begin{equation}
\label{eq:shift_calc}
shift =
\begin{cases}
t_{\text{present}}[23{:}16] - t_{\text{past}}[23{:}16] \\
\quad \text{if } t_{\text{past}}[23{:}16] \leq t_{\text{present}}[23{:}16] \\
t_{\text{present}}[23{:}16] \\
\quad \text{otherwise}
\end{cases}
\end{equation}
}

{\small
\begin{equation}
\label{eq:mem_update}
Mem[\text{ADDR}] =
\begin{cases}
1 + \left( Mem[\text{ADDR}] \gg shift \right) \\
\quad \text{if } shift < 16 \\
1 \\
\quad \text{otherwise}
\end{cases}
\end{equation}
}

{\small
\begin{equation}
\label{eq:mem_shift_subtract}
Mem[\text{ADDR}] =
\begin{cases}
1 + Mem[\text{ADDR}] - shift, \\
\quad \text{if } shift < Mem[\text{ADDR}] \\
1, \\
\quad \text{otherwise}
\end{cases}
\end{equation}
}

\noindent Equation~\ref{eq:shift_calc} calculates the decay term, while Equations~\ref{eq:mem_update} and~\ref{eq:mem_shift_subtract} perform hardware-friendly SETS and SLTS computation respectively.

As illustrated earlier in Fig. \ref{pre_process_overview}, before feeding the data to the interface unit FIFO and display unit FIFO, it is passed through a scaling block where the 16-bit input is converted to 8-bit using scaling and shift operations.

\subsection{Interface Unit}

The interface unit acts as a bridge between the pre-processing and the inference unit. The input data memory organization of the accelerator is different for convolution operations operating over multi-channel and single-channel cases. Initially, the interface controller is in the idle state monitoring the empty signal from the interface FIFO. Once the empty signal is de-asserted, depending upon the model requirements, the controller goes to the data transfer stage. Once the data transfer is complete, the interface unit controller sends the enable signal to the accelerator to begin the operation. The controller remains in the hold state until the done signal is received from the accelerator. After that, it returns to the idle state, monitoring the FIFO empty signal from the interface FIFO and repeating the inference cycle. 

\subsection{RAMAN Accelerator}
The top-level architecture of the RAMAN accelerator is depicted in Fig.~\ref{endtoend_flow}. The global memory (GM) is responsible for storing input event representations, hidden layer activations, and network parameters. To mitigate the overhead associated with frequent global memory access, dedicated activation and parameter caches are employed to exploit temporal data reuse. An activation sparsity engine (ASE) is integrated to reduce computational latency by dynamically bypassing operations involving zero-valued activations. The processing element (PE) array executes multiply-accumulate (MAC) operations, which are subsequently followed by bias addition, ReLU activation, and quantization within a dedicated post-processing (PP) unit. A centralized top-level controller orchestrates the overall dataflow by scheduling operations and dispatching control signals to the constituent modules. %The accelerator architecture is adapted from \cite{krishna2024raman}, and has been redesigned for this purpose. 

\subsection{Data Collection and Training}
The original DVS Gesture dataset was collected through a DVS Camera with 128 $\times$ 128 resolution. Since IMX636 has a resolution of 1280 $\times$ 720 pixels, we collected an in-house dataset using the same 11 gesture classes as the original DVS Gesture dataset to demonstrate the full end-to-end functionality of our pipeline. Five participants were instructed to perform each gesture multiple times in front of an event camera, introducing natural variation in execution speed and style. As part of the preprocessing pipeline, batches of 20K events were converted into 128$\times$128 resolution frame-based representations using SLTS, SETS, standard LTS, ETS, and histogram accumulation, based on the LUT-based approach described in Subsection~\ref{subsec:addr_gen}. These representations were used for both deployment and ablation studies.

The dataset was split into 80:20 training and testing partitions. We obtained 21,932 training frames and 8,197 testing frames from our in-house recordings. For comparison, the original DVS Gesture dataset yielded 5,321 training frames and 1,411 testing frames after preprocessing. We trained two CNNs: a lightweight model termed HOMI-Net16 with $\sim$ 16K parameters and a larger model termed HOMI-Net70 with $\sim$ 70K parameters for dual-channel input (refer to Table \ref{tab:homi_architecture}). Both models accept multi-channel ($N$) inputs of shape $N \times 128 \times 128$ and were trained using quantisation-aware training (QAT) to enable 8-bit deployment.

The models were trained for 1,000 epochs using cross-entropy loss. We used the Adam optimizer with an initial learning rate of $1 \times 10^{-3}$, combined with a cosine annealing schedule for smooth learning rate decay. To improve training robustness, we employed a \textit{top-$k$ loss} strategy, where gradients are backpropagated only through the hardest $k$ samples in each batch. The top-$k$ ratio was progressively reduced using an exponential decay schedule over the training epochs.

\begin{table}[ht]
\centering
\scriptsize
\caption{Layer-wise Architecture Comparison of HOMI-Net16 and HOMI-Net70}
\label{tab:homi_architecture}
\begin{tabularx}{\linewidth}{|c|X|X|}
\hline
\textbf{Layer} & \textbf{HOMI-Net16} & \textbf{HOMI-Net70} \\ \hline
1  & Conv2D (2→16, s=2, k=3, p=1)   & Conv2D (2→16, s=2, k=3, p=1)   \\ \hline
2  & DWConv (16→16, s=2)           & DWConv (16→16, s=1)            \\ \hline
3  & DWConv (16→32, s=2)           & DWConv (16→32, s=2)            \\ \hline
4  & DWConv (32→32, s=2)           & DWConv (32→32, s=1)            \\ \hline
5  & DWConv (32→64, s=1)           & DWConv (32→64, s=2)            \\ \hline
6  & DWConv (64→128, s=2)          & DWConv (64→128, s=1)           \\ \hline
7  & ---                           & DWConv (128→128, s=1)          \\ \hline
8  & ---                           & DWConv (128→256, s=2)          \\ \hline
9  & AdaptiveAvgPool2D (1×1)       & AdaptiveAvgPool2D (1×1)        \\ \hline
10 & Linear (128→11)               & Linear (256→11)                \\ \hline
\end{tabularx}
\vspace{1mm}
\begin{minipage}{0.95\linewidth}
\scriptsize
\textit{Note I:} $s$ = stride, $k$ = kernel size, $p$ = padding, in\_channels $\rightarrow$ out\_channels. \\
\textit{Note II:} DWConv refers to Depthwise Separable Convolution. All convolutional blocks include BatchNorm and ReLU activation.
\end{minipage}
\end{table}

\section{Results}

\subsection{Implementation Results}
This section assesses the performance of HOMI on the Xilinx Ultrascale+MPSoC FPGA. The sensor and PS-PL block, pre-processing module and inference unit operate at different clock rates to extract maximum performance in terms of latency and throughput. The intermediate FIFO modules take care of the data synchronization between multiple clock domains through a double-stage synchronizer configured when the IP was generated.

The implementation of HOMI-Net16, trained on the in-house DVS Gesture dataset, requires a total of 65,746 LUTs, 42,444 FFs, 205 BRAMs, and 14 DSP slices, achieving a low-latency inference time of 1 ms. For the DVS Gesture dataset, the more complex HOMI-Net70 achieves an inference time of 3.593 ms, with a modest increase in resource utilization to accommodate the larger parameter set. The current architecture is optimized for an input resolution of $128 \times 128$, allowing all model parameters and intermediate activations to be stored entirely within the available BRAM resources on the FPGA. For higher spatial resolutions, if the BRAM capacity is exceeded, the on-board DDR memory can be employed for parameter storage and inter-module data transfer, albeit at the cost of increased inference latency.

\subsection{Latency Details}
\begin{figure}[!t]
  \centering
  \includegraphics[width=0.8\columnwidth]{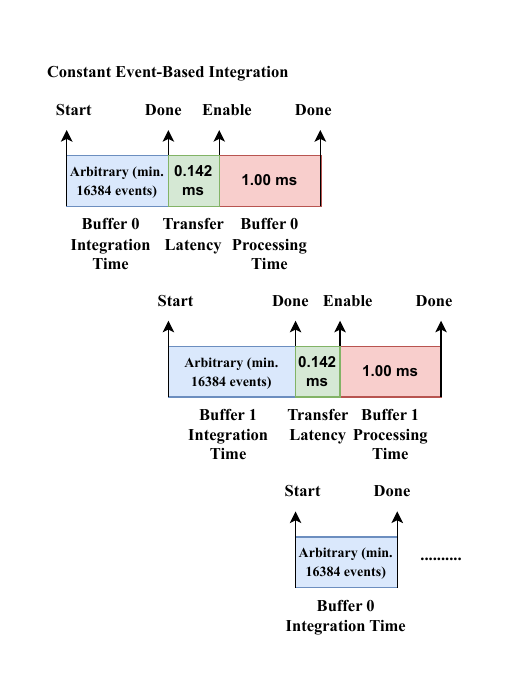} % Use width slightly under 3.5in
  \caption{Latency details for the constant-event based flow. The integration time can be arbitrary, given the asynchronous nature of inputs. Upon receiving the done signal, the integrated event representation is transferred, while the other buffer starts to aggregate events. The transfer latency is 0.142 ms while the inference latency of the classifier is 1ms for HOMI-Net16.}
  \label{timing}
\end{figure}
\par Fig \ref{timing} shows the latency for the end-to-end inference system. In the case of constant time-based integration, the time will be fixed by the counter threshold based on the pre-processing clock. Once a representation is generated, it is transferred to the inference unit via the interface block. Once the data is fed into the inference unit, it operates while the other set of memory starts integration. If the integration latency is greater than the processing latency, then the pipeline will not have any bottleneck for processing; however, if the processing latency is greater than the integration latency, then event dropping can happen. In the case of constant event-based integration, the integration time is variable as it depends on the scene dynamics. Hence, there is a probability of the integration time lesser than the processing time. Hence, in these cases, the limiting factor is the inference system.  

\par In the case of event-based integration, if the probability of such occurrences is less, then it can be handled by increasing the depth of the FIFOs available at the interface unit. The representations will be filled in the FIFO queue and processed once the current inference is completed. In the account of increasing the throughput, given its frugal resource utilization, multiple units of the accelerator can be deployed to achieve the required throughput. For the DVS Gesture implementation on HOMI-Net16, we obtain a processing latency of 1 ms, translating to a throughput of 1000 fps.

\subsection{Ablation Study}

The ablation study presented in Table~\ref{tab:ablation_study} compares different event-based frame representations: ETS, LTS, and histogram accumulation, using a fixed window of 20K events in constant-event mode. Both standard and shift-based variants were evaluated. The shift-based versions consistently outperformed their standard counterparts in terms of accuracy. Among the three representations, ETS consistently achieved the highest accuracy in both shift-based and standard implementations, followed by LTS and histogram accumulation.

A shift factor of 16 was used, corresponding to a temporal decay constant $\tau = \frac{2^{16}}{\ln(2)}$. Using a larger model yielded higher accuracy but at the expense of relatively lower throughput. Interestingly, SETS with 8 input channels, implemented on HOMI-Net16, provides a balanced trade-off by achieving 91.0\% accuracy at 542 fps. The parameter count increases to 19K due to the expanded input dimensionality from additional channels. This demonstrates that with appropriate tuning of both the input encoding and model configuration, it is possible to jointly optimise for accuracy and latency.
These results highlight the trade-off between channel depth, accuracy, and throughput.  Overall, the proposed platform is flexible and can be tuned for either high-accuracy or low-latency operation, or both, depending on the application requirements.

\begin{table}[]
\centering
\scriptsize
\caption{Ablation Study: Shift-Based vs. Standard Frame Generation Methods (ETS, LTS, Histogram) for DVSGesture — Accuracy and Throughput}
\begin{tabular}{|l|l|l|l|l|}
\hline
\begin{tabular}[c]{@{}l@{}}Pre-Processing\\ Method\end{tabular} & \multicolumn{1}{c|}{\begin{tabular}[c]{@{}c@{}}Model with\\ (\# Trainable\\ Parameters)\end{tabular}} & \#N                 & \multicolumn{1}{c|}{\begin{tabular}[c]{@{}c@{}}Through\\ put\\ (fps)\end{tabular}}  & \begin{tabular}[c]{@{}l@{}}Accuracy\\ (\%)\end{tabular}                         \\ \hline
SETS                                                            &                                                                                                                               &                     &                                                                                                            & 94.0  \\ \cline{1-1} \cline{5-5} 
ETS                                                             &                                                                                                                               &                     &                                                                                                            & 90.4                          \\ \cline{1-1} \cline{5-5} 
SLTS                                                            &                                                                                                                               &                     &                                                                                                            & 92.9                          \\ \cline{1-1} \cline{5-5} 
LTS                                                             &                                                                                                                               &                     &                                                                                                            & 92.2                          \\ \cline{1-1} \cline{5-5} 
\begin{tabular}[c]{@{}l@{}}Histogram \\ Accumulation\end{tabular}       & \multirow{-5}{*}{\begin{tabular}[c]{@{}l@{}}HOMI-Net70\\ (70.5K)\end{tabular}}                                               & \multirow{-5}{*}{2} & \multirow{-5}{*}{278}                                                                                      & 91.9                          \\ \hline
SETS                                                            &                                                                                                                               &                     &                                                                                                            & 88.51 \\ \cline{1-1} \cline{5-5} 
ETS                                                             &                                                                                                                               &                     &                                                                                                            & 85.4                          \\ \cline{1-1} \cline{5-5} 
SLTS                                                            &                                                                                                                               &                     &                                                                                                            & 88.2                          \\ \cline{1-1} \cline{5-5} 
LTS                                                             &                                                                                                                               &                     &                                                                                                            & 85.5                          \\ \cline{1-1} \cline{5-5} 
\begin{tabular}[c]{@{}l@{}}Histogram \\ Accumulation\end{tabular}       & \multirow{-5}{*}{\begin{tabular}[c]{@{}l@{}}HOMI-Net16\\ (16.2K)\end{tabular}}                                               & \multirow{-5}{*}{2} & \multirow{-5}{*}{1000}                                                                                      & 88.4                          \\ \hline
SETS                                                            & \begin{tabular}[c]{@{}l@{}}HOMI-Net16\\ (19.9K)\end{tabular}                                                                 & 8                   & 542                                                                                                        & 91.0                          \\ \hline
\end{tabular}
\label{tab:ablation_study}
\vspace{1mm}
\begin{minipage}{0.95\linewidth}
\scriptsize
\textit{Note I:} $N$ = input channels\\
\textit{Note II:} ETS and LTS are the standard Exponential and Linear time surface implementations commonly used.\\
\end{minipage}
\end{table}

\begin{table*}[t]
\centering
\scriptsize
\renewcommand{\arraystretch}{1.2}
\begin{threeparttable}
\caption{Comparison with Prior Hardware Implementations.\vspace{-1mm}}

\begin{tabular}{|>{\centering\arraybackslash}m{1.9cm}|
                >{\centering\arraybackslash}m{1.5cm}|
                >{\centering\arraybackslash}m{1.5cm}|
                >{\centering\arraybackslash}m{1.5cm}|
                >{\centering\arraybackslash}m{1.5cm}|
                >{\centering\arraybackslash}m{1.5cm}|
                >{\centering\arraybackslash}m{1.5cm}|
                >{\centering\arraybackslash}m{1.5cm}|
                >{\centering\arraybackslash}m{1.5cm}|}
\hline
\textbf{Metric} 
& \textbf{\cite{amir2017low}\textsuperscript{\S}} 
& \textbf{\cite{tapiador2020event}} 
& \textbf{\cite{linares2021dynamic}} 
& \textbf{\cite{esda}} 
& \multicolumn{4}{c|}{\textbf{This Work}} \\ \hline
%& & & & & TinyCNN (Ours) & TinyCNN (DVS) & BigCNN (Ours) & BigCNN (DVS) \\ \hline

Pre-Processing & Cascade of Temporal Filters & LTS & Histogram Normalization & - & \multicolumn{4}{c|}{SETS} \\ \hline
Model & SCNN & HOTS & RoShamboNet & MobileNetV2\textsuperscript{\dag} & \multicolumn{2}{c|}{HOMI-Net16} & \multicolumn{2}{c|}{HOMI-Net70} \\ \hline
Dataset & DvsGesture & NavGestures-sit & RoShambo17 & DvsGesture\textsuperscript{\dag} & In-house DVS Gesture & DVS Gesture & In-house DVS Gesture & DVS Gesture \\ \hline
Bit-width & Ternary & 16b & 16b & 8b & 8b & 8b & 8b & 8b \\ \hline
Accuracy (\%) & 94.6 & 93.3 & 99.3 & 93.9 & 97.3 & 88.51 & 98.9 & 94.0 \\ \hline
End-to-End & Yes & Yes & Yes & No & Yes & Yes & Yes & Yes \\ \hline
LUTs & -- & 8.3K\textsuperscript{*} & 266K & 140K & 65.74K (31\%) & 65.74K (31\%)& 71.37K (33\%) & 71.37K (33\%) \\ \hline
FFs & -- & 5.6K\textsuperscript{*} & 139K & 104K & 42.44K (9.89\%) & 42.44K (9.89\%) & 42.53K (9.91\%) & 42.53K (9.91\%) \\ \hline
BRAMs & -- & 18\textsuperscript{*} & 401 & 1134 & 205 (28.7\%) & 205 (28.7\%) & 222 (31.1\%) & 222 (31.1\%) \\ \hline
DSPs & -- & 46\textsuperscript{*} & 657 & 1636 & 14 (0.71\%) & 14 (0.71\%) & 14 (0.71\%) & 14 (0.71\%) \\ \hline
Latency (ms) & 105 & 10\textsuperscript{\ddag} & 8 & 1.19 & 1 & 1 & 3.59 & 3.59 \\ \hline
Throughput (FPS) & -- & 100\textsuperscript{\ddag} & 160 & 839 & 1000 & 1000 & 278 & 278 \\ \hline
\end{tabular}

\vspace{1mm}
\begin{flushleft}
\scriptsize
\textsuperscript{\dag} Multiple implementations of different models on different datasets are available. From the models used for DvsGesture, MobileNetV2 performed better in accuracy, so it is used here for comparison.\\
\textsuperscript{\ddag} Latency reported for event-wise processing. Best case: 0.5~$\mu$s/event, worst case: 6~$\mu$s/event. For 20K events, best-case latency $\approx$10~ms. \\
\textsuperscript{*} Multiple FPGA implementations available; resource-optimized variant shown here.\\
\textsuperscript{\S} SCNN implemented in Samsung 28nm LPP CMOS.
\end{flushleft}

\label{tab:comparison_sota}
\end{threeparttable}
\end{table*}

\subsection{Comparison with Prior Works}
This section compares HOMI with the state-of-the-art (SOA) implementations for end-to-end event-based inference systems. \cite{amir2017low} interfaced the iniLABs DVS128 camera with the NS1e development board that hosts the IBM TrueNorth processor using the AER interface. A cascade of temporal filters was fed to the spiking processor, which implemented a ternarized CNN model as an SNN. From the onset of the Gesture, the overall system latency to classify was 105 ms. Along with the challenge of training SNNs, latency is one of the key metrics where HOMI outperforms. However, it was the first gesture recognition system implemented end-to-end on an event-based system. In \cite{tapiador2020event}, event packets from the computer were sent for processing via the AER interface instead of a DVS. The pre-computed hierarchy of linear time surface (HOTS) templates for different gestures from the NavGestures-sit were compared with the generated event-by-event processed time surface through Euclidean distance. In the best case scenario, an event was processed within 0.5~$\mu$s/event; however, in the worst case, it can go up to 6~$\mu$s/event. Compared to HOMI, the resource utilization is significantly less; however, in terms of latency, even the best-case latency for 20,000 underperforms by almost a factor of 10. In \cite{linares2021dynamic}, an iniLABs DAVIS240C  DVS sensor is interfaced with the Nullhop accelerator via AER. A normalized histogram generated for 2000 events is fed to the accelerator, which performs the inference for five convolutional layers, followed by deploying the fully connected layer on the PS part of the FPGA. HOMI outperforms in terms of both resource utilization and latency. In \cite{esda}, the MobileNetv2 model running on the ESDA architecture outperforms HOMI when compared with the HOMI-Net16 model in terms of accuracy, whilst HOMI outperforms in terms of latency. The HOMI-Net70 model slightly performs better in terms of accuracy, at a relatively lower throughput compared to MobileNetV2 running on the ESDA architecture; however, given the resource utilization metric, HOMI outperforms by utilizing significantly fewer resources, which leaves us an ample amount of room to increase the throughput if the need arises. Also, it is worth mentioning that HOMI is a complete end-to-end deployment, whereas the architecture \cite{esda} does not feature an end-to-end setup. 

\par While most of the end-to-end systems rely on the AER interface, which is highly limited by its data rates, in HOMI, the sensor is directly interfaced with the FPGA via MIPI, which addresses the data rate bottleneck. The pre-processing block also offers flexibility in choosing between various spatiotemporal representations unavailable in the prior implementations. Without any additional overhead of getting the events in ($x$,$y$,$p$,$t$) format, we directly decode the event stream in its native format, which further helps reduce system latency.

\par The accelerator we have adapted is robust enough for deploying different DNN topologies, along with support for quantization and sparsity. For the current implementation, we have mainly relied on the on-chip memory; however, for future requirements, resources are available on the hardware front to support higher-resolution processing and/or deploying multiple classifier units for increased throughput. With the PS-PL integration in place, HOMI opens up various applications in perception systems, bringing a paradigm shift in the next-gen event-based hardware.

\section{Conclusions}

The limitations of conventional frame-based vision systems, particularly in terms of latency, power efficiency, and suitability for edge applications, have motivated the adoption of event-based vision sensors. Event cameras such as the DVS provide sparse, low-latency, and high-temporal-resolution data streams that enable efficient perception under tight computational and energy constraints. These properties make them highly suitable for edge robotics, autonomous systems, and other real-time platforms requiring fast and reliable decision-making.

Custom accelerators deployed on FPGAs are better suited for exploiting the inherent sparse nature of the event data; however, the existing implementations incur high processing latency and/or lack end-to-end systems for performing efficient inference. To address these limitations, we presented HOMI, a fully integrated, end-to-end ultra-fast FPGA-based EdgeAI platform designed for event cameras in a compact form factor. Using the DVSGesture dataset as a representative case study, HOMI achieved an inference latency of 1 ms, demonstrating its effectiveness in high-speed event-based classification.

It is important to emphasize that DVSGesture was selected only as a use case to validate the platform. The underlying architecture is modular and extensible, making it suitable for more complex tasks such as high-speed drone tracking, robotic interaction, or real-time object following. The current resource utilisation on the FPGA is only 33\% of the overall available LUTs, which provides significant room to incorporate more advanced algorithms or expand system capabilities. For instance, the latency of 3.6 ms observed with HOMI-Net70 can be reduced to sub-millisecond by executing multiple HOMI-Net70 instances in parallel, with a corresponding increase in resource usage. To support higher-dimensional input encodings, such as additional input channels, the number of PEs in the accelerator can be scaled up. Beyond CNN inference, additional logic blocks can be added to the pipeline for post-processing tasks such as signal processing, tracking, or decision fusion. The current implementation supports feed-forward networks, and future work will extend the architecture to include temporal models such as Long Short-Term Memory (LSTM) networks to better capture spatio-temporal information in event data.

In conclusion, HOMI offers a flexible, efficient, and scalable end-to-end solution for event-based processing at the edge. Its modular design and low resource footprint allow it to adapt to a wide range of neuromorphic applications while maintaining low latency and high throughput.

\bibliographystyle{IEEEtran} % IEEE style
\bibliography{references}    % your .bib file name (no .bib extension)

\begin{thebibliography}{10}
\providecommand{\url}[1]{#1}
\csname url@samestyle\endcsname
\providecommand{\newblock}{\relax}
\providecommand{\bibinfo}[2]{#2}
\providecommand{\BIBentrySTDinterwordspacing}{\spaceskip=0pt\relax}
\providecommand{\BIBentryALTinterwordstretchfactor}{4}
\providecommand{\BIBentryALTinterwordspacing}{\spaceskip=\fontdimen2\font plus
\BIBentryALTinterwordstretchfactor\fontdimen3\font minus
  \fontdimen4\font\relax}
\providecommand{\BIBforeignlanguage}[2]{{%
\expandafter\ifx\csname l@#1\endcsname\relax
\typeout{** WARNING: IEEEtran.bst: No hyphenation pattern has been}%
\typeout{** loaded for the language `#1'. Using the pattern for}%
\typeout{** the default language instead.}%
\else
\language=\csname l@#1\endcsname
\fi
#2}}
\providecommand{\BIBdecl}{\relax}
\BIBdecl

\bibitem{trinh2018energy}
H.~Trinh, P.~Calyam, D.~Chemodanov, S.~Yao, Q.~Lei, F.~Gao, and K.~Palaniappan,
  ``Energy-aware mobile edge computing and routing for low-latency visual data
  processing,'' \emph{IEEE Transactions on Multimedia}, vol.~20, no.~10, pp.
  2562--2577, 2018.

\bibitem{gallego2020event}
G.~Gallego, T.~Delbr{\"u}ck, G.~Orchard, C.~Bartolozzi, B.~Taba, A.~Censi,
  S.~Leutenegger, A.~J. Davison, J.~Conradt, K.~Daniilidis \emph{et~al.},
  ``Event-based vision: A survey,'' \emph{IEEE Transactions on Pattern Analysis
  and Machine Intelligence}, vol.~44, no.~1, pp. 154--180, 2020.

\bibitem{sengupta2022embedded}
J.~P. Sengupta, M.~Villemur, P.~O. Pouliquen, P.~Julian, and A.~G. Andreou,
  ``Embedded processing pipeline exploration for neuromorphic event based
  perceptual systems,'' in \emph{2022 IEEE International Symposium on Circuits
  and Systems (ISCAS)}.\hskip 1em plus 0.5em minus 0.4em\relax IEEE, 2022, pp.
  486--490.

\bibitem{wang2025towards}
H.~Wang, R.~Guo, P.~Ma, C.~Ruan, X.~Luo, W.~Ding, T.~Zhong, J.~Xu, Y.~Liu, and
  X.~Chen, ``Towards mobile sensing with event cameras on high-agility
  resource-constrained devices: A survey,'' \emph{arXiv preprint
  arXiv:2503.22943}, 2025.

\bibitem{sridharan2024ev}
S.~Sridharan, S.~Selvam, K.~Roy, and A.~Raghunathan, ``Ev-edge: Efficient
  execution of event-based vision algorithms on commodity edge platforms,'' in
  \emph{Proceedings of the 61st ACM/IEEE Design Automation Conference}, 2024,
  pp. 1--6.

\bibitem{kryjak2024event}
T.~Kryjak, ``Event-based vision on fpgas-a survey,'' in \emph{2024 27th
  Euromicro Conference on Digital System Design (DSD)}.\hskip 1em plus 0.5em
  minus 0.4em\relax IEEE, 2024, pp. 541--550.

\bibitem{xu2020case}
C.~Xu, S.~Jiang, G.~Luo, G.~Sun, N.~An, G.~Huang, and X.~Liu, ``The case for
  fpga-based edge computing,'' \emph{IEEE Transactions on Mobile Computing},
  vol.~21, no.~7, pp. 2610--2619, 2020.

\bibitem{yang2024evgnn}
Y.~Yang, A.~Kneip, and C.~Frenkel, ``Evgnn: An event-driven graph neural
  network accelerator for edge vision,'' \emph{IEEE Transactions on Circuits
  and Systems for Artificial Intelligence}, 2024.

\bibitem{esda}
\BIBentryALTinterwordspacing
Y.~Gao, B.~Zhang, Y.~Ding, and H.~K.-H. So, ``A composable dynamic sparse
  dataflow architecture for efficient event-based vision processing on fpga,''
  in \emph{Proceedings of the 2024 ACM/SIGDA International Symposium on Field
  Programmable Gate Arrays}, ser. FPGA '24.\hskip 1em plus 0.5em minus
  0.4em\relax New York, NY, USA: Association for Computing Machinery, 2024, p.
  246–257. [Online]. Available: \url{https://doi.org/10.1145/3626202.3637558}
\BIBentrySTDinterwordspacing

\bibitem{amir2017low}
A.~Amir, B.~Taba, D.~Berg, T.~Melano, J.~McKinstry, C.~Di~Nolfo, T.~Nayak,
  A.~Andreopoulos, G.~Garreau, M.~Mendoza \emph{et~al.}, ``A low power, fully
  event-based gesture recognition system,'' in \emph{Proceedings of the IEEE
  conference on computer vision and pattern recognition}, 2017, pp. 7243--7252.

\bibitem{linares2021dynamic}
A.~Linares-Barranco, A.~Rios-Navarro, S.~Canas-Moreno, E.~Pi{\~n}ero-Fuentes,
  R.~Tapiador-Morales, and T.~Delbruck, ``Dynamic vision sensor integration on
  fpga-based cnn accelerators for high-speed visual classification,'' in
  \emph{International Conference on Neuromorphic Systems 2021}, 2021, pp. 1--7.

\bibitem{tapiador2020event}
R.~Tapiador-Morales, J.-M. Maro, A.~Jimenez-Fernandez, G.~Jimenez-Moreno,
  R.~Benosman, and A.~Linares-Barranco, ``Event-based gesture recognition
  through a hierarchy of time-surfaces for fpga,'' \emph{Sensors}, vol.~20,
  no.~12, p. 3404, 2020.

\bibitem{krishna2024raman}
A.~Krishna, S.~R. Nudurupati, D.~Chandana, P.~Dwivedi, A.~van Schaik,
  M.~Mehendale, and C.~S. Thakur, ``Raman: A re-configurable and sparse tinyml
  accelerator for inference on edge,'' \emph{IEEE Internet of Things Journal},
  2024.

\bibitem{propheseeIMX636}
\BIBentryALTinterwordspacing
{Sony \& Prophesee}, ``Imx636 hd event-based vision sensor,'' Metavision SDK
  Documentation, 2025, 1280×720 resolution, 4.86 µm pixel, $<$100 µs
  latency, $>$120 dB dynamic range, 1 GEPS peak, 50–205 mW power. [Online].
  Available: \url{https://docs.prophesee.ai/stable/hw/sensors/imx636.html}
\BIBentrySTDinterwordspacing

\bibitem{lichtsteiner200564x64}
P.~Lichtsteiner and T.~Delbruck, ``A 64x64 aer logarithmic temporal derivative
  silicon retina,'' in \emph{Research in Microelectronics and Electronics, 2005
  PhD}, vol.~2.\hskip 1em plus 0.5em minus 0.4em\relax IEEE, 2005, pp.
  202--205.

\bibitem{lichtsteiner2006128}
P.~Lichtsteiner, C.~Posch, and T.~Delbruck, ``A 128 x 128 120db 30mw
  asynchronous vision sensor that responds to relative intensity change,'' in
  \emph{2006 IEEE International Solid State Circuits Conference-Digest of
  Technical Papers}.\hskip 1em plus 0.5em minus 0.4em\relax IEEE, 2006, pp.
  2060--2069.

\bibitem{lichtsteiner2008128}
------, ``A 128x128 120 db 15$\mu$s latency asynchronous temporal contrast
  vision sensor,'' \emph{IEEE journal of solid-state circuits}, vol.~43, no.~2,
  pp. 566--576, 2008.

\bibitem{colibriuav}
S.~Bian, L.~Schulthess, G.~Rutishauser, A.~D. Mauro, L.~Benini, and M.~Magno,
  ``Colibriuav: An ultra-fast, energy-efficient neuromorphic edge processing
  uav-platform with event-based and frame-based cameras,'' in \emph{2023 9th
  International Workshop on Advances in Sensors and Interfaces (IWASI)}, 2023,
  pp. 287--292.

\bibitem{swifteagle}
C.~Vogt, M.~Jost, and M.~Magno, ``Swifteagle: An advanced open-source,
  miniaturized fpga uas platform with dual dvs/frame camera for cutting-edge
  low-latency autonomous algorithms,'' in \emph{2024 IEEE/RSJ International
  Conference on Intelligent Robots and Systems (IROS)}, 2024, pp. 8134--8140.

\bibitem{xilinx_mipi}
\BIBentryALTinterwordspacing
{AMD Xilinx}, \emph{{MIPI CSI-2 Receiver Subsystem Product Guide (PG232)}},
  AMD/Xilinx Documentation, 2024, version 5.2, English. [Online]. Available:
  \url{https://docs.amd.com/r/5.2-English/pg232-mipi-csi2-rx/MIPI-CSI-2-Receiver-Subsystem-Product-Guide-PG232}
\BIBentrySTDinterwordspacing

\bibitem{prophesee_evt21}
\BIBentryALTinterwordspacing
{Prophesee S.A.}, \emph{{Prophesee EVT 2.1} encoding format}, Metavision® SDK
  Documentation, 2025, 64‑bit vectorized format, groups events by 32-pixel
  vectors; little‑endian default on GenX320. [Online]. Available:
  \url{https://docs.prophesee.ai/stable/data/encoding_formats/evt21.html}
\BIBentrySTDinterwordspacing

\bibitem{prophesee_evt30}
\BIBentryALTinterwordspacing
------, \emph{{Prophesee EVT 3.0} encoding format}, Metavision® SDK
  Documentation, 2025, 16‑bit compact vector format for high resolution
  sensors; little‑endian default on GenX320. [Online]. Available:
  \url{https://docs.prophesee.ai/stable/data/encoding_formats/evt3.html}
\BIBentrySTDinterwordspacing

\bibitem{mittal2020survey}
S.~Mittal, ``A survey of fpga-based accelerators for convolutional neural
  networks,'' \emph{Neural computing and applications}, vol.~32, no.~4, pp.
  1109--1139, 2020.

\bibitem{davies2018loihi}
M.~Davies, N.~Srinivasa, T.-H. Lin, G.~Chinya, Y.~Cao, S.~H. Choday, G.~Dimou,
  P.~Joshi, N.~Imam, S.~Jain \emph{et~al.}, ``Loihi: A neuromorphic manycore
  processor with on-chip learning,'' \emph{Ieee Micro}, vol.~38, no.~1, pp.
  82--99, 2018.

\bibitem{furber2014spinnaker}
S.~B. Furber, F.~Galluppi, S.~Temple, and L.~A. Plana, ``The spinnaker
  project,'' \emph{Proceedings of the IEEE}, vol. 102, no.~5, pp. 652--665,
  2014.

\bibitem{bouvier2019spiking}
M.~Bouvier, A.~Valentian, T.~Mesquida, F.~Rummens, M.~Reyboz, E.~Vianello, and
  E.~Beigne, ``Spiking neural networks hardware implementations and challenges:
  A survey,'' \emph{ACM Journal on Emerging Technologies in Computing Systems
  (JETC)}, vol.~15, no.~2, pp. 1--35, 2019.

\bibitem{nullhop}
\BIBentryALTinterwordspacing
A.~Aimar, H.~Mostafa, E.~Calabrese, A.~Rios-Navarro, R.~Tapiador-Morales, I.-A.
  Lungu, M.~B. Milde, F.~Corradi, A.~Linares-Barranco, S.-C. Liu, and
  T.~Delbruck, ``Nullhop: A flexible convolutional neural network accelerator
  based on sparse representations of feature maps,'' \emph{IEEE Transactions on
  Neural Networks and Learning Systems}, vol.~30, no.~3, p. 644–656, Mar.
  2019. [Online]. Available: \url{http://dx.doi.org/10.1109/TNNLS.2018.2852335}
\BIBentrySTDinterwordspacing

\bibitem{bonazzifpgadrone}
\BIBentryALTinterwordspacing
P.~Bonazzi, C.~Vogt, M.~Jost, L.~Khacef, F.~Paredes-Vallés, and M.~Magno,
  ``Towards low-latency event-based obstacle avoidance on a fpga-drone,'' 2025.
  [Online]. Available: \url{https://arxiv.org/abs/2504.10400}
\BIBentrySTDinterwordspacing

\bibitem{lagorce2016hots}
X.~Lagorce, G.~Orchard, F.~Galluppi, B.~E. Shi, and R.~B. Benosman, ``Hots: a
  hierarchy of event-based time-surfaces for pattern recognition,'' \emph{IEEE
  transactions on pattern analysis and machine intelligence}, vol.~39, no.~7,
  pp. 1346--1359, 2016.

\bibitem{sironi2018hats}
A.~Sironi, M.~Brambilla, N.~Bourdis, X.~Lagorce, and R.~Benosman, ``Hats:
  Histograms of averaged time surfaces for robust event-based object
  classification,'' in \emph{Proceedings of the IEEE conference on computer
  vision and pattern recognition}, 2018, pp. 1731--1740.

\bibitem{blachut2023high}
K.~Blachut and T.~Kryjak, ``High-definition event frame generation using soc
  fpga devices,'' in \emph{2023 Signal Processing: Algorithms, Architectures,
  Arrangements, and Applications (SPA)}.\hskip 1em plus 0.5em minus 0.4em\relax
  IEEE, 2023, pp. 106--111.

\bibitem{prophesee_genx320}
\BIBentryALTinterwordspacing
{Prophesee S.A.}, \emph{{Prophesee GenX320} event-based sensor}, Product
  datasheet / SDK documentation, 2025, 320~$\times$~320 resolution, ultra-low
  latency (36~µW), embedded features and supports EVT~2.0, EVT~2.1 (default)
  and EVT~3.0 (MP version). [Online]. Available:
  \url{https://www.prophesee.ai/event-based-sensor-genx320/}
\BIBentrySTDinterwordspacing

\end{thebibliography}
\end{document}